\documentclass[prd,superscriptaddress,twocolumn, 
preprintnumbers,showpacs,nofootinbib,10pt]{revtex4-1} 
\pdfoutput=1
\usepackage{amssymb,amsmath,graphicx,epsfig} 
\usepackage{hyperref,color,float,caption,array}
\usepackage{float}
\usepackage{graphicx}
\hypersetup{linktocpage=true}
\hypersetup{
colorlinks=true,   
linkcolor=blue,    
citecolor=blue,    
filecolor=blue,    
urlcolor=blue      
}
\newcommand{\gsim}{\lower.7ex\hbox{$\;\stackrel{\textstyle>}{\sim}\;$}}
\newcommand{\lsim}{\lower.7ex\hbox{$\;\stackrel{\textstyle<}{\sim}\;$}}
\newcommand{\db}{\bar{d}}

\def\beq{\begin{equation}}
\def\eeq{\end{equation}}
\def\bea{\begin{eqnarray}}
\def\eea{\end{eqnarray}}
\def\bmat{\begin{pmatrix}}
\def\emat{\end{pmatrix}}
\def\bei{\begin{itemize}}
\def\eei{\end{itemize}}

\makeatletter
\renewcommand{\section}{\@startsection{section}{1}{0em}%
        {-3.25ex \@plus -1ex \@minus -.2ex}%
        {2.0ex \@plus.2ex}%
        {\normalfont\large\bfseries}}
\renewcommand{\subsection}{\@startsection{subsection}{2}{0em}%
        {-2.75ex\@plus -1ex \@minus -.2ex}%
        {1.25ex \@plus .2ex}%
        {\normalfont\bfseries}}
\renewcommand{\subsubsection}%
        {\@startsection{subsubsection}{3}{0em}%
        {-2.0ex\@plus -1ex \@minus -.2ex}%
        {1.0ex \@plus .2ex}%
        {\normalfont\itshape}}
\makeatother
\begin{document} 
\title{Is the Higgs Mechanism of Fermion Mass Generation a Fact?
\\ 
A Yukawa-less First-Two-Generation Model}
\author{Diptimoy Ghosh}
\affiliation{\normalfont{Department of Particle Physics and Astrophysics, 
Weizmann Institute of Science, Rehovot 76100, Israel}}
\author{Rick Sandeepan Gupta}
\affiliation{\normalfont{Department of Particle Physics and Astrophysics, 
Weizmann Institute of Science, Rehovot 76100, Israel}}
\author{Gilad Perez} 
\affiliation{\normalfont{Department of Particle Physics and Astrophysics, 
Weizmann Institute of Science, Rehovot 76100, Israel}} 
\begin{abstract} 
It is now established that the major source of electroweak symmetry breaking (EWSB) 
is due to the observed Higgs particle. However, whether the Higgs mechanism is responsible 
for the generation of all the fermion masses, in particular, the fermions of the first two 
generations, is an open question. In this letter we present a construction where the light 
fermion masses are generated through a secondary, subdominant and sequestered source of 
EWSB. This fits well with the approximate U(2) global symmetry of the observed structure 
of the flavor sector.
We first realise the above idea using a calculable two Higgs doublet model. 
We then show that the first two generation masses could come from technicolor dynamics, 
while the third generation fermions, as well as the electroweak gauge bosons get their masses 
dominantly from the Higgs mechanism. We also discuss how the small CKM mixing between the first 
two generations and the third generation, and soft mixing between the sequestered EWSB
components arise in this setup. A typical prediction of this scenario 
is a significant reduction of the couplings of the observed Higgs boson to the first two 
generation of fermions. 
\end{abstract}
\preprint{}
\maketitle

The mechanism of Electroweak Symmetry Breaking (EWSB) and the generation of 
fermion masses are among the main frontiers of research in high energy physics. 
The discovery of a Higgs-like particle and the 
measurements of its couplings to the Standard Model (SM) gauge bosons have 
established that the Higgs mechanism is indeed the correct picture for it, at 
least to leading order. However, whether the Higgs 
particle is also responsible for all the fermion masses is an open 
question. We have direct indications that
the observed Higgs particle couples to the third generation 
fermions with roughly SM strengths
\bea \label{eq:3rdavg} 
\hspace{-0.2cm}
\mu_{t\bar t h,  \,b,\, \tau} = 2.2\pm0.6, \,0.71\pm0.31, \,0.97\pm0.23\,, \eea
where we have averaged the ATLAS~\cite{ATLAS-CONF-2014-011,Aad:2014xzb,Aad:2015vsa} and 
CMS~\cite{Khachatryan:2014qaa,Chatrchyan:2013zna,Chatrchyan:2014nva} results for the 
corresponding signal strengths.  
The couplings to the first two generation of fermions (also referred as light fermions below) 
have not been measured and are only weakly constrained:
~\cite{Aad:2014xva,Khachatryan:2014aep,Perez:2015lra,Perez:2015aoa} 
\begin{equation} \label{eq:muemu}
\mu_{\mu}\leq 7 \, , \,
\mu_{e}\leq 4 \times 10^5 \, , \, \mu_{c}\leq 180\,,
\end{equation}
while for the $u,d,s$ quarks no direct bounds exist at present 
(see~\cite{Kagan:2014ila,Bodwin:2013gca,Brivio:2015fxa,Koenig:2015pha,
Altmannshofer:2015qra,Aad:2015sda,Bodwin:2014bpa} for related discussions).
In addition, the observed flavor sector consists of a large hierarchy between the 
first two and the third generation fermion masses and the mixing angles. This is 
related to the celebrated SM flavor puzzle that is consistent with an approximate 
pattern of U(3)/U(2) symmetry breaking~\cite{D'Ambrosio:2002ex,Kagan:2009bn}.

Below we explore the possibility that the above approximate U(2) structure 
is linked with the way EWSB is communicated to the flavor sector.
We propose that two approximately sequestered sources of EWSB exist in 
nature~\cite{GiladAspenTalk}. 
The major one, is due to a SM-like Higgs doublet, $\Phi_3$, that couples predominantly 
to the third generation fermions. An additional subdominant source, 
$\Phi_{12}$, couples mostly to the first two generations and induces their masses. 
The second source of EWSB $\Phi_{12}$ has very little to do with the observed Higgs particle. 
It could be a cousin of the Higgs or could arise due to 
strong-dynamics, and not be associated with any weakly coupled physics. 
More concretely, we assume that $\Phi_3$  has a mass of about 125 GeV with the 
following couplings
\bea
{\cal L}_{3}^Y &\supset& -\overline{Q}^{(i)}_L 
[ Y_{3~ij}^d \, \Phi_3 \, d_R^{(j)} +  Y_{3~ij}^u \, \widetilde{\Phi}_3 \, u_R^{(j)}]
+ \rm h.c. \nonumber \label{LH}
\eea
with $Y^{u,d}_{3~ij}\approx y^{t,b}\delta_{33}$. Assuming that 
the other source of EWSB transforms as a doublet (to preserve custodial symmetry), 
$\Phi_{12}$, the  mass terms  of the light fermions can be effectively written as, 
\bea
{\cal L}_{12}^Y &\supset& -\overline{Q}^{(i)}_L 
[ Y_{12~ij}^d \, \Phi_{12} \, d_R^{(j)} +  Y_{12~ij}^u \, \widetilde{\Phi}_{12} \, u_R^{(j)}]+
\rm h.c. \nonumber \label{LH2}
\eea
where $i,j$ run from 1 to 2 and,
\bea
v^2_{12}=\langle \Phi_{12} \rangle^2  \ll  v_3^2= \langle \Phi_{3} \rangle^2\,, 
\ \ v^2_{12}+v_3^2 = v^2\,, 
\eea
with $v = 246~{\rm GeV}.$
Later, we will explore the possibility that $\Phi_{12}$ is actually a condensate, 
$\langle \bar{Q}Q\rangle$,  of fermions of a technicolor sector.  If the sector 
containing $\Phi_3$ and the third generation of fermions is completely decoupled 
from the second source of EWSB, the observed Higgs boson would have no couplings to 
the light fermions.   

Two conceptual problems arise, however,  if the two EWSB sectors are completely 
sequestered. First, the (13) and (23) elements of the Cabibbo-Kobayashi-Maskawa (CKM) 
matrix are not generated. 
In order to generate these CKM entries, we will assume that $\Phi_3$ couples very 
weakly to the light fermions,
\bea
Y^{u,d}_{3~ij}= y^{t,b}\delta_{33}+\epsilon^{u,d}_{ij} \,, 
\label{eps}
\eea
where, as shown below, the strength of the $\epsilon_{ij}$ couplings are dictated by the 
observed CKM (13) and (23) elements.
The second issue is that, if there is no coupling between the two sectors at all, 
we can define two different global SU(2) symmetries for $\Phi_{12}$ and $\Phi_3$.
As a result, when $\Phi_3$ and $\Phi_{12}$ develop Vacuum Expectation Values (VEVs), 
both the SU(2) symmetries are broken and six Goldstone bosons arise in the low 
energy spectrum. Three of these Goldstone bosons are absorbed by $W_L^{\pm}$ and 
$Z_L$, once the vectorial combination of the two SU(2) global symmetries is gauged. 
As the gauging explicitly breaks the axial combination of the two SU(2) global 
symmetries, the remaining three Goldstone bosons get small masses at the loop level. 
However, these masses are too small to evade collider constraints and a further 
source of explicit breaking is required to raise their masses. For this, we add, 
in our scalar potential, a $\mu$-term,
\bea
\mu \, \Phi_3^\dagger \, \Phi_{12}+{\rm h.c.}\,,
\label{muterm}
\eea
that breaks this symmetry softly, and lifts the Goldstone boson masses. Note that 
at the spurionic level, once we have the couplings $\epsilon_{ij}$ in Eq.~\eqref{eps} 
the coupling, $\mu$, is also allowed. This means that generically, a mechanism that generates 
the $\epsilon_{ij}$ would also induce the $\mu$-term. In our technicolor 
construction below, this is indeed 
the case.

With the addition of the term in Eq.~\eqref{eps} the light fermions get masses from 
two EWSB sources, thus violating a ``Natural Flavor Conservation" principle~\cite{Glashow:1976nt}. 
This is expected to lead to tree level Flavor Changing Neutral Current (FCNC) processes, suppressed by 
powers of $\epsilon_{ij}$ (see for eg., \cite{Gupta:2009wn}). We will see that, in order to generate CKM 
elements with the correct magnitude, the size of
$\epsilon_{ij}$ needed is small enough such 
that FCNCs are within experimental bounds. In this sense our model is similar in spirit to 
other models  that violate NFC
Ref.~\cite{Blechman:2010cs,Buras:2010mh,Jung:2010ik,Dery:2013aba,Branco:2011iw,Kiers:1998ry,Botella:2009pq}  but 
nevertheless satisfies flavor bounds. 

In order to understand these issues more concretely, we now consider an effective 
2HDM toy model for $\Phi_3$ and $\Phi_{12}$, which will allow us to understand the relevant phenomenological constraints without  committing to a specific UV completion.  The following scalar potential is consistent with the setup described above ~\cite{foot},
\bea
&& V(\Phi_{12},\Phi_3) = 
\mu^2_1 |\Phi_{12}|^2 + \mu_2^2 |\Phi_3|^2 + \mu (\Phi_{12}^\dagger \Phi_3 + {\rm h.c.}) 
\nonumber \\
&& \hspace{2cm} + \lambda_1 |\Phi_{12}|^4 + \lambda_2 |\Phi_3|^4 \, .
\label{pot}
\eea 
The couplings $\mu_1$, $\mu_2$, $\lambda_1$, 
$\lambda_2$ are real due to hermiticity of the Lagrangian. Moreover, the phase of $\mu$ 
can be absorbed in $\Phi_3$. We have in the spectrum a pseudoscalar Higgs $A$ and two charged 
Higgs bosons $H^{\pm}$ which model the pseudo goldstone bosons discussed above 
Eq.~\eqref{muterm}. There are also two CP-even scalars $h$ and $H$. The masses and the 
mixing angle, $\beta$, for the pseudoscalar and charged states are given by,
$m^2_{H^\pm} = m^2_{A^0}= - \mu v^2 / v_{12} v_3, \, \, \tan \beta =v_{3}/v_{12}$ while the masses and the mixing angle, $\alpha$, for 
the CP-even states are given by,
\bea
m^{2}_{h,\,H}  &=&  \lambda_1 v_{12}^2 + \lambda_2 v_{3}^2 -  \frac{\mu v^2}{2 \, v_{12} \, v_3}  \nonumber  \\
&\mp & \sqrt{ \left(\lambda_1 v_{12}^2 - \lambda_2 v_{3}^2 + 
\frac{\mu(v^2_{12}-v^2_3)}{2\, v_{12} \, v_3}\right)^2 + \mu^2} \,, \nonumber  \\
\tan 2\alpha 
&=&-\frac{2 m_{A^0}^2 s_\beta c_\beta}{\sqrt{(m_H^2-m_h^2)^2-4 m_{A^0}^4 s_\beta^2 c_\beta^2}} \,,
\eea
where $c_\theta \equiv \cos \theta$ and $s_\theta \equiv \sin \theta$.
The coupling of the quarks with the Higgs bosons can be written as, 
\bea
\hspace*{-.63cm}{\cal L}_Y^{(q)} \hspace*{-.1cm}&\supset& -\dfrac{h}{v}\bigg\{\bar{u} \bigg[(s_{\beta-\alpha}M_u + c_{\beta-\alpha}Y_u )P_R + \nonumber \\
&& \hspace{1cm}(s_{\beta-\alpha}M_u + c_{\beta-\alpha}Y_u^\dagger)P_L\bigg]  u  + u \leftrightarrow d \bigg\} \nonumber \\
&-& \bigg(h \to H, s_{\beta-\alpha} \to c_{\beta-\alpha}, 
c_{\beta-\alpha} \to - s_{\beta-\alpha} \bigg)\nonumber \\
&+&\dfrac{i A}{v}\bigg[\bar{u} (Y_u P_R - Y_u^\dagger P_L) u - 
\bar{d} (Y_d P_R - Y_d^\dagger P_L) d\bigg] \nonumber \\
&+&\bigg[\dfrac{H^+}{v}  \bar{u}(Y_u^\dagger V_{\rm CKM} P_L - V_{\rm CKM}Y_d P_R) d + {\rm h.c.} \bigg], \label{bigeq}
\eea
where the matrices $Y_u$ and $Y_d$ are defined as,
\bea
Y_{u, d} = V_L^{u,d \, \dagger} \big(-v_{3} Y_{12}^{u,d}  + v_{12} Y_{3}^{u,d} \big) V_R^{u,d} / \sqrt{2} \, ,
\label{ydyu}
\eea
and the matrices $V_L^{u}$, $V_L^{d}$, $V_R^{u}$ and $V_R^{d}$ are defined through the 
following equations,
\bea
M_{u,d} = V_L^{u,d \, \dagger} (v_{12} Y_{12}^{u,d}  + v_{3} Y_{3}^{u,d} ) V_R^{u,d} /\sqrt{2} \, .
\label{massmtr}
\eea
Here, $M_u$ and $M_d$ are the (diagonal) mass matrices.  
As for $Y_{12}^{u,d}$ and $Y_{3}^{u,d}$, without loss of generality we can assume that 
$Y_{12}^{d}$ has the form,
\bea
Y_{12}^{d}&=&
{\rm diag}(y_d,y_{s},0) \, 
\eea
where $y_{d,s}=\sqrt{2}m_{d,s}/v_{12}$ upto ${\cal O}(\epsilon^2)$ corrections. 
In the same basis,
\bea
Y_{3}^{d}&=&\left( 
\begin{array}{ccc}
{\cal O}(\epsilon^{d*}_{L1} \epsilon^{d}_{R1})&{\cal O}(\epsilon^{d*}_{L1}\epsilon^{d}_{R2})&\epsilon^{d*}_{L1} \\
{\cal O}(\epsilon^{d*}_{L2}\epsilon^{d}_{R1})&{\cal O}(\epsilon^{d*}_{L2}\epsilon^{d}_{R2})&\epsilon^{d*}_{L2}\\
\epsilon^{d}_{R1} &\epsilon^{d}_{R2}&y_{b}
\end{array}
\right) \, .
\label{y3d}
\eea
where $y_{b}=\sqrt{2}m_{b}/v_{3}$. In the above basis, $Y^u_{12}$ is misaligned from the up sector mass basis by an 
angle ${\cal O}(V_{\rm CKM}^{12})$. In the basis where $Y^u_{12} =
{\rm diag}(y_u,y_{c},0)$ is diagonal we define $Y_{3}^{u}$ as in Eq.~\eqref{y3d} but 
with the replacement $d \to u$ and $y_b \to y_t$. We take 
$\epsilon_{Li}^{u} \sim \epsilon_{Li}^{d}$\,.

Assuming the above form, we diagonalise $M_{u,d}$ (see Eq.~\eqref{massmtr}) and 
obtain $V^{u,d}_L$ and $V^{u,d}_R$. Then using $V_{\rm CKM}=V^{u \dagger}_L V^{d}_L$ 
we find that  the $\epsilon_{Li}^{u,d}$ are related to the CKM matrix elements as follows,
\bea
\epsilon^{u}_{L2} \sim \epsilon^{d}_{L2} &=& \hat{y}_b V_{\rm CKM}^{23}/s_\beta \, , \nonumber \\
\epsilon^{u}_{L1}\sim \epsilon^{d}_{L1}&=&\hat{y}_b V_{\rm CKM}^{13}/s_\beta-\hat{y}_b V_{\rm CKM}^{12} V_{\rm CKM}^{23}/s_\beta
\eea
where $\hat{y}_f = \sqrt{2} m_f/v$. Note that we still have freedom to choose 
$\epsilon_R^{u,d}$. Using Eq.~\eqref{ydyu} we find that the matrix $Y_q$, which governs 
the tree level FCNC interactions of the pseudoscalar and the charged Higgs bosons, to be
\bea
Y_q &=& \cot \beta \, M_q -  \\
&& \hspace{-1.2cm}\frac{v}{\sqrt{2} c_\beta}\left( 
\begin{array}{ccc}
{\cal O}(\epsilon^4) &\epsilon^{q}_{L1} \epsilon^{q*}_{R2} s_\beta/\hat{y}_3 & s^2_\beta\epsilon^{q}_{L1}\epsilon^{q*~2}_{R2}/\hat{y}^2_3\\
\epsilon^{q*}_{L2} \epsilon^{q*}_{R1} s_\beta/\hat{y}_3&\hat{y}_2/s_\beta&\epsilon^{q*}_{R2}\hat{y}_2/\hat{y}_3\\\
 s^2_\beta\epsilon^{q*}_{R1}\epsilon^{q~2}_{L2}/\hat{y}^2_3&\epsilon^{q}_{L2}\hat{y}_2/\hat{y}_3&s_\beta \hat{y}_2  \epsilon^{q}_{L2}\epsilon^{q*}_{R2}/\hat{y}^2_3\
\end{array}
\right)\nonumber
\eea
where $q=u,d$, $\hat{y}_3=\hat{y}_{t,b}$ and $\hat{y}_2=\hat{y}_{c,s}$. 
As expected $Y_{q}$ has $\epsilon_{ij}$ suppressed off-diagonal entries which lead to 
Higgs mediated FCNCs. For example, the $\Delta F =2$ operators 
$Q_2 = \db^{\alpha}_R  q^{\alpha}_L \db^{\beta}_R q^{\beta}_L \, ,  
{\tilde Q}_2 = \db^{\alpha}_L  q^{\alpha}_R \db^{\beta}_L q^{\beta}_R \, ,
Q_4  =  \db^{\alpha}_R  b^{\alpha}_L \db^{\beta}_L b^{\beta}_R$ are generated by tree 
level exchange of the $h \, ,H$ and $A$. 
While the constraints from $B -\bar{B}$ mixing are trivially satisfied (due to 
the approximate alignment  between $Y_{d}$ and $M_{d}$ which is only broken 
by $y_{s}$ effects), 
contribution to $\Delta M_K$ is just below the experimental sensitivity 
for $\epsilon_{Li} \sim \epsilon_{Ri}$. Reducing the ratio $r^{\epsilon}_{RL}\equiv \epsilon_{Ri}/\epsilon_{Li}$ 
one can further suppress the contribution to Kaon mixing. If ${\cal O}(1)$ phases are present then the 
bound from $\epsilon_K$ requires 
$r^{\epsilon}_{RL}\lesssim 0.1\,$. 

Note that we could have also generated  the (13) and (23) elements of the CKM matrix by adding a perturbation, 
$\epsilon_{ij}$ to $Y_{12}^{u,d}$, instead of $Y_3^{u,d}$. In this case, however, it turns out that contributions to  
$B -\bar{B}$ mixing are a bit larger than the allowed values. This is because here, unlike  the previous case,  
the approximate alignment  between $Y_{d}$ and $M_{d}$ for $y_{s}\to 0$ is not present.
\begin{figure*}[t]
\begin{tabular}{cc}
\includegraphics[scale=0.5]{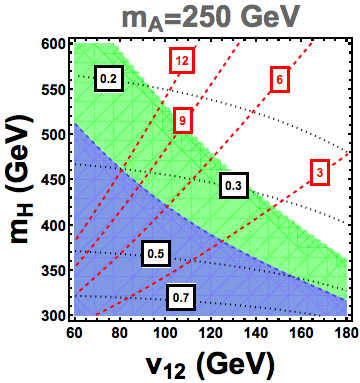} \ \ & \ \ 
\includegraphics[scale=0.5]{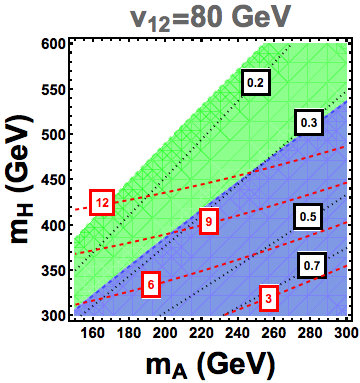}
\end{tabular}
\caption{In blue we show the region allowed by Electroweak precision constraints 
at 95\% CL considering only the 2HDM contribution. The green region shows the improvement 
when an additional UV contribution \{S,T\} =\{0.25,0.25\} is added. The red 
and black dashed lines are contours of fixed $\lambda_1$ and fixed $\kappa_f$ respectively. 
Note that, as the signal strength $\mu_f = \kappa_f^2$, the reduction in the signal strength is more significant.
\label{fig1}}
\end{figure*}

Let us now discuss constraints from Electroweak Precision Observables (EWPO) and other 
phenomenological implications. Following the expressions given in \cite{Branco:2011iw}, 
we compute the Peskin-Takeuchi parameters $S$, $T$ and $U$ \cite{Peskin:1990zt,Peskin:1991sw} 
and show the allowed region in Fig.~\ref{fig1}. The constraints 
from EWPO give an upper bound on $v_{12}$. For example, taking $m_H=450$ and $m_A= 250$ GeV 
we find the bound $v_{12} < 80{\rm \, GeV}$. To evade the direct constraints from the Drell-Yan 
production of $H^\pm$ at LEP one also requires $m_{H^+} \gtrsim 100~{\rm GeV}$, which, in turn, 
provides a lower bound of $\mu \gtrsim 50 - 80~{\rm GeV}$ for $v_{12} \approx 80 - 160~{\rm GeV}$. 
Note that, while the couplings of the observed Higgs boson to the electroweak gauge bosons see a reduction 
of ${\cal O}(v_{12}^2/v^2)$ with respect to the SM value, couplings to the third generation fermions get an 
enhancement of the same order. Thus we need precision Higgs measurements to reach a ${\cal O}(10 \%)$ sensitivity to detect these deviations.

Using Eq.~\eqref{bigeq} we also find that $\kappa_f$, the coupling of the observed Higgs boson 
to a light fermion $f$ normalised to its SM value $m_f/v$, can be written as, 
\bea
\label{kappa_f}
\kappa_f = -\frac{s_\alpha}{c_\beta}+{\cal O}(\epsilon^2)
\hspace{0.5cm}\longrightarrow^{^{\hspace{-0.9cm} m_A \ll m_H}} \, 
s_\beta \frac{m_A^2}{m_H^2}.
\eea
Hence, for $m_H^2 \gg m_A^2$ these couplings can be considerably reduced with respect to the 
SM. In order to  quantify the extent of reduction in the Higgs coupling to the light 
fermions in our 2HDM framework, in Fig.~\ref{fig1} we show (in blue) the allowed region in the 
$m_A - m_H$ plane (for fixed $v_{12}$) and the $v_{12} - m_H$ plane (for fixed $m_A$) 
overlayed with contours of $\kappa_f$. One can notice that 
a factor of  $2-3$ reduction in the coupling is possible. This is,  however, at the cost of large values of one of the 
quartic couplings, $\lambda_1$ ($\lambda_2$ is almost fixed by the observed 
Higgs boson mass). 

Thus, in the region of the parameter space, where the observed Higgs couplings to the light 
fermions are reduced, a large value of $\lambda_1$ renders the theory nonperturbative at the 
TeV scale. In this limit, our setup finds a natural  embedding in a strongly coupled theory 
like technicolor where the VEV of $\Phi_{12}$ is identified with $\sqrt{N_D} F_T$, $F_T$ 
and $N_D$ being the techni-pion decay constant and the number of TC doublets respectively. 
The scalars $A^0$ and $H^\pm$ are identified with the (pseudo) goldstone bosons of the 
TC chiral symmetry breaking. The mass of the heavy CP-even Higgs, $m_H$, should be identified 
with the scale of the resonances, 
\bea
m_{\rho_T} \sim (F_T/f_\pi) \sqrt{3/N_T}  \, m_\rho
& \sim & 500~{\rm GeV} \sqrt{\frac{6}{N_T N_D}} \nonumber
\eea 
The above estimate has been done by scaling QCD and taking 
$v_{12} = 100~{\rm GeV} \, , \, f_\pi = 125~{\rm MeV}$.

Note that, in order to generate the Yukawa matrix $Y_{12}^{u,d}$, the SM and 
TC fermions have to be embedded in larger multiplets that transform under the so called 
extended technicolor (ETC) group. The ETC gauge group breaks into SM $\otimes$ TC group 
at the ETC scale $M_{\rm ETC}$ . Integrating out the gauge bosons corresponding to the 
broken generators introduces four-fermion operators like,
\bea
(g_{\rm ETC}^2/M_{\rm ETC}^2)  \, ({\bar q}  q)({\bar Q}  Q)
\eea
where $q$ and $Q$ are SM and TC quarks respectively,  $M_{\rm ETC}$ is the mass 
of the gauge boson and $g_{\rm ETC}$ is the gauge coupling.
This operator generates a mass term for $q$ when, at a lower scale, the TC group becomes 
strongly coupled and the quark $Q$ forms a condensate,
\bea
m_q \sim (g_{\rm ETC}^2/M_{\rm ETC}^2)  \, \langle\overline{Q}Q\rangle_{\rm ETC}.
\eea
$\langle \overline{Q}Q\rangle_{\rm ETC}$ is the value of the technifermion condensate 
at the ETC scale which can be related to its value at the TC scale by,
\bea
\langle \overline{Q}Q\rangle_{\rm ETC} = \langle \overline{Q}Q\rangle_{\rm TC} {\rm Exp}\bigg[
\int_{\Lambda_{\rm TC}}^{M_{\rm ETC}} d\mu/\mu \, \gamma_m(\mu)\bigg]. \nonumber
\eea
where $\langle \overline{Q}Q\rangle_{\rm TC}=4 \pi F_T^3$ and $\gamma$ is the anomalous 
dimension of the operator.   Note that the third generation fermions are also promoted to ETC multiplets, 
but the ETC breaking pattern is such that, like the first generation fermions, they get small masses from 
the TC sector, thus generating  $Y^{u,d}_{12}$ of the form we have assumed.  In this setup, therefore,  the lightness of the first two generation fermions is a natural consequence of the fact that the dominant contribution to their masses is from  higher dimensional  operators generated at a high scale.

A nice feature of this technicolor setup is that  the all important $\mu$ coupling is generated automatically from the 
yukawa couplings, $Y_{3}^{u,d}$,  of $\Phi_3$. In order to couple $\Phi_3$ 
to SM fermions, $\Phi_3 \bar{q}q$ in an ETC invariant way,  we must also  couple it to the condensate 
$\Phi_3 \langle\bar{Q}Q \rangle$, thus generating a $\mu$-term in the potential. The top Yukawa coupling, for instance, can generate,
\bea
\mu \sim  \frac{4 \pi v_{12}^2}{N_D^{3/2}}\sim \frac{(150~{\rm GeV})^2}{(N_D/3)^{3/2}}
\eea
for $v_{12}\sim 100$ GeV which gives $m_A \sim 250~{\rm GeV}$. 
One can also check that, in this setup, quartic terms in the  scalar potential 
in Eq.~\eqref{pot} would arise from irrelevant operators above the TC scale and would thus be 
subdominant compared to the $\mu$-term.

Let us now discuss in some detail  how the usual phenomenological 
problems associated with technicolor theories can be evaded in our case. Possibly the  main issues with technicolor models are:
(1) EW precision tests and 
(2) FCNCs. The flavor problem of standard 
technicolor theories is far less severe in our case, because we do not demand that the 
condensate accounts for the large top mass whereas the tension with the EWPO 
is ameliorated because of the presence of a light Higgs in our theory.  
Let us discuss these one by one. 

The tension with flavour physics arises because the  extended technicolor 
interactions are also expected to generate four fermion interactions involving  only 
SM fermions which give rise to FCNC interactions, for example, those contributing 
to $\Delta M_K$.  One can estimate the $\Delta S =2$ effective Lagrangian to be,  
\bea
(g_{\rm ETC}^2/M_{\rm ETC}^2)  \, \theta_{sd}^2 ({\bar s} \gamma_\mu P_{L,R} d)
({\bar s} \gamma^\mu P_{L,R} d)
\eea
where, $\theta_{sd}$ is the mixing angle governing the $s \to d$ transition and 
should be of the order of Cabibbo angle. 
Using the experimental value of $\Delta M_K$~\cite{Agashe:2014kda,foot1} one gets the bound, 
\bea
&& g_{\rm ETC} \sqrt{{\rm Re}(\theta_{sd}^2)}/(M_{\rm ETC}) \lesssim (700\rm ~TeV)^{-1} \,, \, 
\rm which~gives,  \nonumber \\
&& \hspace{-0.0cm} m_q \lesssim \left(M_{\rm ETC}/\Lambda_{\rm TC}\right)^{\bar{\gamma}}
\frac{4 \pi F_T^3}{(700 \rm \,TeV)^2 {\rm Re}(\theta_{sd}^2)}
\eea
where,
$\bar{\gamma} =\int_{\Lambda_{\rm TC}}^{M_{\rm ETC}}\frac{d\mu}{\mu}\gamma_m(\mu) \, / \, \log(M_{\rm ETC}/\Lambda_{\rm TC})$
is a parameter which depends on the UV details of the strongly coupled sector. Theories with $\bar{\gamma}\approx 0$ are called running TC  theories whereas $\bar{\gamma} \to 1$ is called the walking limit. Taking $\bar{\gamma}=1$, $F_{\rm TC} \sim 100~{\rm GeV}$  and 
$\theta_{sd} \sim 0.1$ we get $ m_q \lesssim  2.5~{\rm GeV}$ for 
$M_{\rm ETC}/\Lambda_{\rm TC} \sim 1000$.
Hence, obtaining the correct charm quark mass is not a problem in the the walking limit. It is also clear that getting the correct top mass would have been impossible, if we did not have an additional Higgs doublet. Values smaller than $\bar{\gamma}=1$ can be made compatible with data, if flavour symmetries are imposed in the ETC sector \cite{Hill:2002ap} so that smaller values of $M_{\rm ETC}$ are allowed. Walking also helps raise the mass of the  pseudogoldstone bosons from chiral symmetry breaking (other than those corresponding to $A,H^\pm$) to the TeV 
scale~\cite{Hill:2002ap}.

We now discuss the constraints from EWPO on the contribution of the technicolor resonances. The contribution of the resonances  to the Peskin-Takeuchi $S$-parameter 
is known to be positive and the estimate in \cite{Peskin:1990zt,Peskin:1991sw} gives,
$
S\sim 0.25 \, N_T N_D/6 \,,$
which assumes that the technicolor theory is a scaled up version of QCD. The usual 
tension of technicolor theories with EWPO, however, is much milder in our case 
because we already have a light Higgs in our spectrum. The electroweak fit assuming 
a 125 GeV Higgs gives at 95$\%$ C.L. \cite{Baak:2014ora,Ciuchini:2013pca},
$
S = 0.05 \pm 0.22 \,, \, 
T = 0.09 \pm 0.26 \,.
$
We thus see that $S=0.25$  is still allowed by the fit at 95$\%$ C.L., although a 
contribution to the $T$ parameter of similar magnitude is also required. This can be 
achieved in various ways and we refer to \cite{Hill:2002ap} and the references therein for more 
details. In fact, adding an additional contribution $\Delta S \,,  \Delta T \sim 0.25$ to the 
infrared  contribution enlarges the allowed parameter space, as can be seen from Fig.~\ref{fig1}.
Note that the absence of a light Higgs boson shifts the allowed region 
in the $S-T$ plane by $\Delta S \sim -0.15$ and $\Delta T \sim 0.20$ \cite{Beringer:1900zz} 
making the tension with EWPO much stronger. Going back to Eq.~\eqref{kappa_f}, we can now see 
that in the TC limit the reduction in the Couplings of the Higgs bosons to the light 
fermions can be even more significant than the 2HDM case. For example, from Fig.~\ref{fig1} 
we get the following estimates for the signal strengths,
\bea
\label{reduction}
\big(\mu_{f}\big)_{\rm 2HDM,\, \rm TC}\, \gtrsim\  0.09 \,, \ 0.03 \,.
\eea

Note that in the framework presented above the fact that $v_{12}$ and $v_3$ are of the same order seemingly gives rise to a coincidence problem. 
In a more complete model, however, it is possible that  that the spontaneous symmetry breaking in the SM-like Higgs sector would trigger the 
dynamical breaking in the Technicolor-like sector which would make this whole setup more compelling.

In summary, in this letter we construct a model where the masses of the fermions of the first two generations  
are not associated with the Standard Model Higgs mechanism, but with another subdominant sequestered 
source of electroweak symmetry breaking e.g., a technicolor sector. 
This structure can naturally explain the smallness of the mixing between the light and heavy generations as well 
as potentially the lightness of the first two generations.  Furthermore, in the technicolor realisation the smallness 
of the masses is a natural consequence of the fact that they are induced by irrelevant operators.  We find that a 
reduction of the  couplings of the observed Higgs boson to the light fermions is a generic prediction of this setup.  
Our framework can be experimentally tested by direct detection of the pseudogoldstone states $A, H^\pm$ and the TC resonances. 
We also predict ${\cal O}(10\%)$ deviations  of Higgs couplings to gauge bosons 
and third generation fermions  from  Standard Model values  which can be measured in future precision Higgs precision  measurements. 
A direct measurement of the   couplings of the observed Higgs  to light fermions will be possible at  the ILC and, in particular,  with even 
greater sensitivity at the  TLEP~\cite{Perez:2015lra}.

{\bf Acknowledgment}
GP is supported by the IRG, ISF, and ERC-2013-CoG grant (TOPCHARM \# 614794). 
We thank Eilam Gross, Yevgeny Kats, Leandro Da Rold and Yotam Soreq for useful 
discussions.   

{\bf Note added}
While this paper was in the final stage of completion, ref~\cite{Altmannshofer:2015esa} 
appeared on arXiv with a related proposal, however, with a different motivation and a limited
overlap.


\begin{thebibliography}{39}%
\makeatletter
\providecommand \@ifxundefined [1]{%
 \@ifx{#1\undefined}
}%
\providecommand \@ifnum [1]{%
 \ifnum #1\expandafter \@firstoftwo
 \else \expandafter \@secondoftwo
 \fi
}%
\providecommand \@ifx [1]{%
 \ifx #1\expandafter \@firstoftwo
 \else \expandafter \@secondoftwo
 \fi
}%
\providecommand \natexlab [1]{#1}%
\providecommand \enquote  [1]{``#1''}%
\providecommand \bibnamefont  [1]{#1}%
\providecommand \bibfnamefont [1]{#1}%
\providecommand \citenamefont [1]{#1}%
\providecommand \href@noop [0]{\@secondoftwo}%
\providecommand \href [0]{\begingroup \@sanitize@url \@href}%
\providecommand \@href[1]{\@@startlink{#1}\@@href}%
\providecommand \@@href[1]{\endgroup#1\@@endlink}%
\providecommand \@sanitize@url [0]{\catcode `\\12\catcode `\$12\catcode
  `\&12\catcode `\#12\catcode `\^12\catcode `\_12\catcode `\%12\relax}%
\providecommand \@@startlink[1]{}%
\providecommand \@@endlink[0]{}%
\providecommand \url  [0]{\begingroup\@sanitize@url \@url }%
\providecommand \@url [1]{\endgroup\@href {#1}{\urlprefix }}%
\providecommand \urlprefix  [0]{URL }%
\providecommand \Eprint [0]{\href }%
\providecommand \doibase [0]{http://dx.doi.org/}%
\providecommand \selectlanguage [0]{\@gobble}%
\providecommand \bibinfo  [0]{\@secondoftwo}%
\providecommand \bibfield  [0]{\@secondoftwo}%
\providecommand \translation [1]{[#1]}%
\providecommand \BibitemOpen [0]{}%
\providecommand \bibitemStop [0]{}%
\providecommand \bibitemNoStop [0]{.\EOS\space}%
\providecommand \EOS [0]{\spacefactor3000\relax}%
\providecommand \BibitemShut  [1]{\csname bibitem#1\endcsname}%
\let\auto@bib@innerbib\@empty
\bibitem [{ATL()}]{ATLAS-CONF-2014-011}%
  \BibitemOpen
  \href@noop {} {\bibinfo  {journal} {The ATLAS Collaboration,
  ATLAS-CONF-2014-011}\ }\BibitemShut {NoStop}%
\bibitem [{\citenamefont {Aad}\ \emph {et~al.}(2015{\natexlab{a}})\citenamefont
  {Aad} \emph {et~al.}}]{Aad:2014xzb}%
  \BibitemOpen
\bibfield  {journal} {  }\bibfield  {author} {\bibinfo {author} {\bibfnamefont
  {G.}~\bibnamefont {Aad}} \emph {et~al.} (\bibinfo {collaboration} {ATLAS}),\
  }\href {\doibase 10.1007/JHEP01(2015)069} {\bibfield  {journal} {\bibinfo
  {journal} {JHEP}\ }\textbf {\bibinfo {volume} {01}},\ \bibinfo {pages} {069}
  (\bibinfo {year} {2015}{\natexlab{a}})},\ \Eprint
  {http://arxiv.org/abs/1409.6212} {arXiv:1409.6212 [hep-ex]} \BibitemShut
  {NoStop}%
\bibitem [{\citenamefont {Aad}\ \emph {et~al.}(2015{\natexlab{b}})\citenamefont
  {Aad} \emph {et~al.}}]{Aad:2015vsa}%
  \BibitemOpen
  \bibfield  {author} {\bibinfo {author} {\bibfnamefont {G.}~\bibnamefont
  {Aad}} \emph {et~al.} (\bibinfo {collaboration} {ATLAS}),\ }\href {\doibase
  10.1007/JHEP04(2015)117} {\bibfield  {journal} {\bibinfo  {journal} {JHEP}\
  }\textbf {\bibinfo {volume} {04}},\ \bibinfo {pages} {117} (\bibinfo {year}
  {2015}{\natexlab{b}})},\ \Eprint {http://arxiv.org/abs/1501.04943}
  {arXiv:1501.04943 [hep-ex]} \BibitemShut {NoStop}%
\bibitem [{\citenamefont {Khachatryan}\ \emph {et~al.}(2014)\citenamefont
  {Khachatryan} \emph {et~al.}}]{Khachatryan:2014qaa}%
  \BibitemOpen
  \bibfield  {author} {\bibinfo {author} {\bibfnamefont {V.}~\bibnamefont
  {Khachatryan}} \emph {et~al.} (\bibinfo {collaboration} {CMS}),\ }\href
  {\doibase 10.1007/JHEP09(2014)087, 10.1007/JHEP10(2014)106} {\bibfield
  {journal} {\bibinfo  {journal} {JHEP}\ }\textbf {\bibinfo {volume} {09}},\
  \bibinfo {pages} {087} (\bibinfo {year} {2014})},\ \bibinfo {note} {[Erratum:
  JHEP10,106(2014)]},\ \Eprint {http://arxiv.org/abs/1408.1682}
  {arXiv:1408.1682 [hep-ex]} \BibitemShut {NoStop}%
\bibitem [{\citenamefont {Chatrchyan}\ \emph
  {et~al.}(2014{\natexlab{a}})\citenamefont {Chatrchyan} \emph
  {et~al.}}]{Chatrchyan:2013zna}%
  \BibitemOpen
  \bibfield  {author} {\bibinfo {author} {\bibfnamefont {S.}~\bibnamefont
  {Chatrchyan}} \emph {et~al.} (\bibinfo {collaboration} {CMS}),\ }\href
  {\doibase 10.1103/PhysRevD.89.012003} {\bibfield  {journal} {\bibinfo
  {journal} {Phys. Rev.}\ }\textbf {\bibinfo {volume} {D89}},\ \bibinfo {pages}
  {012003} (\bibinfo {year} {2014}{\natexlab{a}})},\ \Eprint
  {http://arxiv.org/abs/1310.3687} {arXiv:1310.3687 [hep-ex]} \BibitemShut
  {NoStop}%
\bibitem [{\citenamefont {Chatrchyan}\ \emph
  {et~al.}(2014{\natexlab{b}})\citenamefont {Chatrchyan} \emph
  {et~al.}}]{Chatrchyan:2014nva}%
  \BibitemOpen
  \bibfield  {author} {\bibinfo {author} {\bibfnamefont {S.}~\bibnamefont
  {Chatrchyan}} \emph {et~al.} (\bibinfo {collaboration} {CMS}),\ }\href
  {\doibase 10.1007/JHEP05(2014)104} {\bibfield  {journal} {\bibinfo  {journal}
  {JHEP}\ }\textbf {\bibinfo {volume} {05}},\ \bibinfo {pages} {104} (\bibinfo
  {year} {2014}{\natexlab{b}})},\ \Eprint {http://arxiv.org/abs/1401.5041}
  {arXiv:1401.5041 [hep-ex]} \BibitemShut {NoStop}%
\bibitem [{\citenamefont {Aad}\ \emph {et~al.}(2014)\citenamefont {Aad} \emph
  {et~al.}}]{Aad:2014xva}%
  \BibitemOpen
  \bibfield  {author} {\bibinfo {author} {\bibfnamefont {G.}~\bibnamefont
  {Aad}} \emph {et~al.} (\bibinfo {collaboration} {ATLAS}),\ }\href {\doibase
  10.1016/j.physletb.2014.09.008} {\bibfield  {journal} {\bibinfo  {journal}
  {Phys. Lett.}\ }\textbf {\bibinfo {volume} {B738}},\ \bibinfo {pages} {68}
  (\bibinfo {year} {2014})},\ \Eprint {http://arxiv.org/abs/1406.7663}
  {arXiv:1406.7663 [hep-ex]} \BibitemShut {NoStop}%
\bibitem [{\citenamefont {Khachatryan}\ \emph {et~al.}(2015)\citenamefont
  {Khachatryan} \emph {et~al.}}]{Khachatryan:2014aep}%
  \BibitemOpen
  \bibfield  {author} {\bibinfo {author} {\bibfnamefont {V.}~\bibnamefont
  {Khachatryan}} \emph {et~al.} (\bibinfo {collaboration} {CMS}),\ }\href
  {\doibase 10.1016/j.physletb.2015.03.048} {\bibfield  {journal} {\bibinfo
  {journal} {Phys. Lett.}\ }\textbf {\bibinfo {volume} {B744}},\ \bibinfo
  {pages} {184} (\bibinfo {year} {2015})},\ \Eprint
  {http://arxiv.org/abs/1410.6679} {arXiv:1410.6679 [hep-ex]} \BibitemShut
  {NoStop}%
\bibitem [{\citenamefont {Perez}\ \emph
  {et~al.}(2015{\natexlab{a}})\citenamefont {Perez}, \citenamefont {Soreq},
  \citenamefont {Stamou},\ and\ \citenamefont {Tobioka}}]{Perez:2015lra}%
  \BibitemOpen
  \bibfield  {author} {\bibinfo {author} {\bibfnamefont {G.}~\bibnamefont
  {Perez}}, \bibinfo {author} {\bibfnamefont {Y.}~\bibnamefont {Soreq}},
  \bibinfo {author} {\bibfnamefont {E.}~\bibnamefont {Stamou}}, \ and\ \bibinfo
  {author} {\bibfnamefont {K.}~\bibnamefont {Tobioka}},\ }\href@noop {} {\
  (\bibinfo {year} {2015}{\natexlab{a}})},\ \Eprint
  {http://arxiv.org/abs/1505.06689} {arXiv:1505.06689 [hep-ph]} \BibitemShut
  {NoStop}%
\bibitem [{\citenamefont {Perez}\ \emph
  {et~al.}(2015{\natexlab{b}})\citenamefont {Perez}, \citenamefont {Soreq},
  \citenamefont {Stamou},\ and\ \citenamefont {Tobioka}}]{Perez:2015aoa}%
  \BibitemOpen
  \bibfield  {author} {\bibinfo {author} {\bibfnamefont {G.}~\bibnamefont
  {Perez}}, \bibinfo {author} {\bibfnamefont {Y.}~\bibnamefont {Soreq}},
  \bibinfo {author} {\bibfnamefont {E.}~\bibnamefont {Stamou}}, \ and\ \bibinfo
  {author} {\bibfnamefont {K.}~\bibnamefont {Tobioka}},\ }\href@noop {} {\
  (\bibinfo {year} {2015}{\natexlab{b}})},\ \Eprint
  {http://arxiv.org/abs/1503.00290} {arXiv:1503.00290 [hep-ph]} \BibitemShut
  {NoStop}%
\bibitem [{\citenamefont {Kagan}\ \emph {et~al.}(2015)\citenamefont {Kagan},
  \citenamefont {Perez}, \citenamefont {Petriello}, \citenamefont {Soreq},
  \citenamefont {Stoynev},\ and\ \citenamefont {Zupan}}]{Kagan:2014ila}%
  \BibitemOpen
  \bibfield  {author} {\bibinfo {author} {\bibfnamefont {A.~L.}\ \bibnamefont
  {Kagan}}, \bibinfo {author} {\bibfnamefont {G.}~\bibnamefont {Perez}},
  \bibinfo {author} {\bibfnamefont {F.}~\bibnamefont {Petriello}}, \bibinfo
  {author} {\bibfnamefont {Y.}~\bibnamefont {Soreq}}, \bibinfo {author}
  {\bibfnamefont {S.}~\bibnamefont {Stoynev}}, \ and\ \bibinfo {author}
  {\bibfnamefont {J.}~\bibnamefont {Zupan}},\ }\href {\doibase
  10.1103/PhysRevLett.114.101802} {\bibfield  {journal} {\bibinfo  {journal}
  {Phys. Rev. Lett.}\ }\textbf {\bibinfo {volume} {114}},\ \bibinfo {pages}
  {101802} (\bibinfo {year} {2015})},\ \Eprint {http://arxiv.org/abs/1406.1722}
  {arXiv:1406.1722 [hep-ph]} \BibitemShut {NoStop}%
\bibitem [{\citenamefont {Bodwin}\ \emph {et~al.}(2013)\citenamefont {Bodwin},
  \citenamefont {Petriello}, \citenamefont {Stoynev},\ and\ \citenamefont
  {Velasco}}]{Bodwin:2013gca}%
  \BibitemOpen
  \bibfield  {author} {\bibinfo {author} {\bibfnamefont {G.~T.}\ \bibnamefont
  {Bodwin}}, \bibinfo {author} {\bibfnamefont {F.}~\bibnamefont {Petriello}},
  \bibinfo {author} {\bibfnamefont {S.}~\bibnamefont {Stoynev}}, \ and\
  \bibinfo {author} {\bibfnamefont {M.}~\bibnamefont {Velasco}},\ }\href
  {\doibase 10.1103/PhysRevD.88.053003} {\bibfield  {journal} {\bibinfo
  {journal} {Phys.Rev.}\ }\textbf {\bibinfo {volume} {D88}},\ \bibinfo {pages}
  {053003} (\bibinfo {year} {2013})},\ \Eprint {http://arxiv.org/abs/1306.5770}
  {arXiv:1306.5770 [hep-ph]} \BibitemShut {NoStop}%
\bibitem [{\citenamefont {Brivio}\ \emph {et~al.}(2015)\citenamefont {Brivio},
  \citenamefont {Goertz},\ and\ \citenamefont {Isidori}}]{Brivio:2015fxa}%
  \BibitemOpen
  \bibfield  {author} {\bibinfo {author} {\bibfnamefont {I.}~\bibnamefont
  {Brivio}}, \bibinfo {author} {\bibfnamefont {F.}~\bibnamefont {Goertz}}, \
  and\ \bibinfo {author} {\bibfnamefont {G.}~\bibnamefont {Isidori}},\
  }\href@noop {} {\  (\bibinfo {year} {2015})},\ \Eprint
  {http://arxiv.org/abs/1507.02916} {arXiv:1507.02916 [hep-ph]} \BibitemShut
  {NoStop}%
\bibitem [{\citenamefont {Koenig}\ and\ \citenamefont
  {Neubert}(2015)}]{Koenig:2015pha}%
  \BibitemOpen
  \bibfield  {author} {\bibinfo {author} {\bibfnamefont {M.}~\bibnamefont
  {Koenig}}\ and\ \bibinfo {author} {\bibfnamefont {M.}~\bibnamefont
  {Neubert}},\ }\href@noop {} {\  (\bibinfo {year} {2015})},\ \Eprint
  {http://arxiv.org/abs/1505.03870} {arXiv:1505.03870 [hep-ph]} \BibitemShut
  {NoStop}%
\bibitem [{\citenamefont {Altmannshofer}\ \emph
  {et~al.}(2015{\natexlab{a}})\citenamefont {Altmannshofer}, \citenamefont
  {Brod},\ and\ \citenamefont {Schmaltz}}]{Altmannshofer:2015qra}%
  \BibitemOpen
  \bibfield  {author} {\bibinfo {author} {\bibfnamefont {W.}~\bibnamefont
  {Altmannshofer}}, \bibinfo {author} {\bibfnamefont {J.}~\bibnamefont {Brod}},
  \ and\ \bibinfo {author} {\bibfnamefont {M.}~\bibnamefont {Schmaltz}},\
  }\href {\doibase 10.1007/JHEP05(2015)125} {\bibfield  {journal} {\bibinfo
  {journal} {JHEP}\ }\textbf {\bibinfo {volume} {05}},\ \bibinfo {pages} {125}
  (\bibinfo {year} {2015}{\natexlab{a}})},\ \Eprint
  {http://arxiv.org/abs/1503.04830} {arXiv:1503.04830 [hep-ph]} \BibitemShut
  {NoStop}%
\bibitem [{\citenamefont {Aad}\ \emph {et~al.}(2015{\natexlab{c}})\citenamefont
  {Aad} \emph {et~al.}}]{Aad:2015sda}%
  \BibitemOpen
  \bibfield  {author} {\bibinfo {author} {\bibfnamefont {G.}~\bibnamefont
  {Aad}} \emph {et~al.} (\bibinfo {collaboration} {ATLAS}),\ }\href {\doibase
  10.1103/PhysRevLett.114.121801} {\bibfield  {journal} {\bibinfo  {journal}
  {Phys. Rev. Lett.}\ }\textbf {\bibinfo {volume} {114}},\ \bibinfo {pages}
  {121801} (\bibinfo {year} {2015}{\natexlab{c}})},\ \Eprint
  {http://arxiv.org/abs/1501.03276} {arXiv:1501.03276 [hep-ex]} \BibitemShut
  {NoStop}%
\bibitem [{\citenamefont {Bodwin}\ \emph {et~al.}(2014)\citenamefont {Bodwin},
  \citenamefont {Chung}, \citenamefont {Ee}, \citenamefont {Lee},\ and\
  \citenamefont {Petriello}}]{Bodwin:2014bpa}%
  \BibitemOpen
  \bibfield  {author} {\bibinfo {author} {\bibfnamefont {G.~T.}\ \bibnamefont
  {Bodwin}}, \bibinfo {author} {\bibfnamefont {H.~S.}\ \bibnamefont {Chung}},
  \bibinfo {author} {\bibfnamefont {J.-H.}\ \bibnamefont {Ee}}, \bibinfo
  {author} {\bibfnamefont {J.}~\bibnamefont {Lee}}, \ and\ \bibinfo {author}
  {\bibfnamefont {F.}~\bibnamefont {Petriello}},\ }\href {\doibase
  10.1103/PhysRevD.90.113010} {\bibfield  {journal} {\bibinfo  {journal} {Phys.
  Rev.}\ }\textbf {\bibinfo {volume} {D90}},\ \bibinfo {pages} {113010}
  (\bibinfo {year} {2014})},\ \Eprint {http://arxiv.org/abs/1407.6695}
  {arXiv:1407.6695 [hep-ph]} \BibitemShut {NoStop}%
\bibitem [{\citenamefont {D'Ambrosio}\ \emph {et~al.}(2002)\citenamefont
  {D'Ambrosio}, \citenamefont {Giudice}, \citenamefont {Isidori},\ and\
  \citenamefont {Strumia}}]{D'Ambrosio:2002ex}%
  \BibitemOpen
  \bibfield  {author} {\bibinfo {author} {\bibfnamefont {G.}~\bibnamefont
  {D'Ambrosio}}, \bibinfo {author} {\bibfnamefont {G.~F.}\ \bibnamefont
  {Giudice}}, \bibinfo {author} {\bibfnamefont {G.}~\bibnamefont {Isidori}}, \
  and\ \bibinfo {author} {\bibfnamefont {A.}~\bibnamefont {Strumia}},\ }\href
  {\doibase 10.1016/S0550-3213(02)00836-2} {\bibfield  {journal} {\bibinfo
  {journal} {Nucl. Phys.}\ }\textbf {\bibinfo {volume} {B645}},\ \bibinfo
  {pages} {155} (\bibinfo {year} {2002})},\ \Eprint
  {http://arxiv.org/abs/hep-ph/0207036} {arXiv:hep-ph/0207036 [hep-ph]}
  \BibitemShut {NoStop}%
\bibitem [{\citenamefont {Kagan}\ \emph {et~al.}(2009)\citenamefont {Kagan},
  \citenamefont {Perez}, \citenamefont {Volansky},\ and\ \citenamefont
  {Zupan}}]{Kagan:2009bn}%
  \BibitemOpen
  \bibfield  {author} {\bibinfo {author} {\bibfnamefont {A.~L.}\ \bibnamefont
  {Kagan}}, \bibinfo {author} {\bibfnamefont {G.}~\bibnamefont {Perez}},
  \bibinfo {author} {\bibfnamefont {T.}~\bibnamefont {Volansky}}, \ and\
  \bibinfo {author} {\bibfnamefont {J.}~\bibnamefont {Zupan}},\ }\href
  {\doibase 10.1103/PhysRevD.80.076002} {\bibfield  {journal} {\bibinfo
  {journal} {Phys. Rev.}\ }\textbf {\bibinfo {volume} {D80}},\ \bibinfo {pages}
  {076002} (\bibinfo {year} {2009})},\ \Eprint {http://arxiv.org/abs/0903.1794}
  {arXiv:0903.1794 [hep-ph]} \BibitemShut {NoStop}%
\bibitem [{\citenamefont {Perez}()}]{GiladAspenTalk}%
  \BibitemOpen
  \bibfield  {author} {\bibinfo {author} {\bibfnamefont {G.}~\bibnamefont
  {Perez}},\ }\href@noop {} {\bibinfo  {journal} {Talk given at the Exploring
  Phys. with FCC, Aspen Winter Conference, Jan/2015}\ }\BibitemShut {NoStop}%
\bibitem [{\citenamefont {Glashow}\ and\ \citenamefont
  {Weinberg}(1977)}]{Glashow:1976nt}%
  \BibitemOpen
\bibfield  {journal} {  }\bibfield  {author} {\bibinfo {author} {\bibfnamefont
  {S.~L.}\ \bibnamefont {Glashow}}\ and\ \bibinfo {author} {\bibfnamefont
  {S.}~\bibnamefont {Weinberg}},\ }\href {\doibase 10.1103/PhysRevD.15.1958}
  {\bibfield  {journal} {\bibinfo  {journal} {Phys. Rev.}\ }\textbf {\bibinfo
  {volume} {D15}},\ \bibinfo {pages} {1958} (\bibinfo {year}
  {1977})}\BibitemShut {NoStop}%
\bibitem [{\citenamefont {Gupta}\ and\ \citenamefont
  {Wells}(2010)}]{Gupta:2009wn}%
  \BibitemOpen
  \bibfield  {author} {\bibinfo {author} {\bibfnamefont {R.~S.}\ \bibnamefont
  {Gupta}}\ and\ \bibinfo {author} {\bibfnamefont {J.~D.}\ \bibnamefont
  {Wells}},\ }\href {\doibase 10.1103/PhysRevD.81.055012} {\bibfield  {journal}
  {\bibinfo  {journal} {Phys. Rev.}\ }\textbf {\bibinfo {volume} {D81}},\
  \bibinfo {pages} {055012} (\bibinfo {year} {2010})},\ \Eprint
  {http://arxiv.org/abs/0912.0267} {arXiv:0912.0267 [hep-ph]} \BibitemShut
  {NoStop}%
\bibitem [{\citenamefont {Blechman}\ \emph {et~al.}(2010)\citenamefont
  {Blechman}, \citenamefont {Petrov},\ and\ \citenamefont
  {Yeghiyan}}]{Blechman:2010cs}%
  \BibitemOpen
  \bibfield  {author} {\bibinfo {author} {\bibfnamefont {A.~E.}\ \bibnamefont
  {Blechman}}, \bibinfo {author} {\bibfnamefont {A.~A.}\ \bibnamefont
  {Petrov}}, \ and\ \bibinfo {author} {\bibfnamefont {G.}~\bibnamefont
  {Yeghiyan}},\ }\href {\doibase 10.1007/JHEP11(2010)075} {\bibfield  {journal}
  {\bibinfo  {journal} {JHEP}\ }\textbf {\bibinfo {volume} {11}},\ \bibinfo
  {pages} {075} (\bibinfo {year} {2010})},\ \Eprint
  {http://arxiv.org/abs/1009.1612} {arXiv:1009.1612 [hep-ph]} \BibitemShut
  {NoStop}%
\bibitem [{\citenamefont {Buras}\ \emph {et~al.}(2010)\citenamefont {Buras},
  \citenamefont {Carlucci}, \citenamefont {Gori},\ and\ \citenamefont
  {Isidori}}]{Buras:2010mh}%
  \BibitemOpen
  \bibfield  {author} {\bibinfo {author} {\bibfnamefont {A.~J.}\ \bibnamefont
  {Buras}}, \bibinfo {author} {\bibfnamefont {M.~V.}\ \bibnamefont {Carlucci}},
  \bibinfo {author} {\bibfnamefont {S.}~\bibnamefont {Gori}}, \ and\ \bibinfo
  {author} {\bibfnamefont {G.}~\bibnamefont {Isidori}},\ }\href {\doibase
  10.1007/JHEP10(2010)009} {\bibfield  {journal} {\bibinfo  {journal} {JHEP}\
  }\textbf {\bibinfo {volume} {10}},\ \bibinfo {pages} {009} (\bibinfo {year}
  {2010})},\ \Eprint {http://arxiv.org/abs/1005.5310} {arXiv:1005.5310
  [hep-ph]} \BibitemShut {NoStop}%
\bibitem [{\citenamefont {Jung}\ \emph {et~al.}(2010)\citenamefont {Jung},
  \citenamefont {Pich},\ and\ \citenamefont {Tuzon}}]{Jung:2010ik}%
  \BibitemOpen
  \bibfield  {author} {\bibinfo {author} {\bibfnamefont {M.}~\bibnamefont
  {Jung}}, \bibinfo {author} {\bibfnamefont {A.}~\bibnamefont {Pich}}, \ and\
  \bibinfo {author} {\bibfnamefont {P.}~\bibnamefont {Tuzon}},\ }\href
  {\doibase 10.1007/JHEP11(2010)003} {\bibfield  {journal} {\bibinfo  {journal}
  {JHEP}\ }\textbf {\bibinfo {volume} {11}},\ \bibinfo {pages} {003} (\bibinfo
  {year} {2010})},\ \Eprint {http://arxiv.org/abs/1006.0470} {arXiv:1006.0470
  [hep-ph]} \BibitemShut {NoStop}%
\bibitem [{\citenamefont {Dery}\ \emph {et~al.}(2013)\citenamefont {Dery},
  \citenamefont {Efrati}, \citenamefont {Hiller}, \citenamefont {Hochberg},\
  and\ \citenamefont {Nir}}]{Dery:2013aba}%
  \BibitemOpen
  \bibfield  {author} {\bibinfo {author} {\bibfnamefont {A.}~\bibnamefont
  {Dery}}, \bibinfo {author} {\bibfnamefont {A.}~\bibnamefont {Efrati}},
  \bibinfo {author} {\bibfnamefont {G.}~\bibnamefont {Hiller}}, \bibinfo
  {author} {\bibfnamefont {Y.}~\bibnamefont {Hochberg}}, \ and\ \bibinfo
  {author} {\bibfnamefont {Y.}~\bibnamefont {Nir}},\ }\href {\doibase
  10.1007/JHEP08(2013)006} {\bibfield  {journal} {\bibinfo  {journal} {JHEP}\
  }\textbf {\bibinfo {volume} {1308}},\ \bibinfo {pages} {006} (\bibinfo {year}
  {2013})},\ \Eprint {http://arxiv.org/abs/1304.6727} {arXiv:1304.6727
  [hep-ph]} \BibitemShut {NoStop}%
\bibitem [{\citenamefont {Branco}\ \emph {et~al.}(2012)\citenamefont {Branco},
  \citenamefont {Ferreira}, \citenamefont {Lavoura}, \citenamefont {Rebelo},
  \citenamefont {Sher},\ and\ \citenamefont {Silva}}]{Branco:2011iw}%
  \BibitemOpen
  \bibfield  {author} {\bibinfo {author} {\bibfnamefont {G.~C.}\ \bibnamefont
  {Branco}}, \bibinfo {author} {\bibfnamefont {P.~M.}\ \bibnamefont
  {Ferreira}}, \bibinfo {author} {\bibfnamefont {L.}~\bibnamefont {Lavoura}},
  \bibinfo {author} {\bibfnamefont {M.~N.}\ \bibnamefont {Rebelo}}, \bibinfo
  {author} {\bibfnamefont {M.}~\bibnamefont {Sher}}, \ and\ \bibinfo {author}
  {\bibfnamefont {J.~P.}\ \bibnamefont {Silva}},\ }\href {\doibase
  10.1016/j.physrep.2012.02.002} {\bibfield  {journal} {\bibinfo  {journal}
  {Phys. Rept.}\ }\textbf {\bibinfo {volume} {516}},\ \bibinfo {pages} {1}
  (\bibinfo {year} {2012})},\ \Eprint {http://arxiv.org/abs/1106.0034}
  {arXiv:1106.0034 [hep-ph]} \BibitemShut {NoStop}%
\bibitem [{\citenamefont {Kiers}\ \emph {et~al.}(1999)\citenamefont {Kiers},
  \citenamefont {Soni},\ and\ \citenamefont {Wu}}]{Kiers:1998ry}%
  \BibitemOpen
  \bibfield  {author} {\bibinfo {author} {\bibfnamefont {K.}~\bibnamefont
  {Kiers}}, \bibinfo {author} {\bibfnamefont {A.}~\bibnamefont {Soni}}, \ and\
  \bibinfo {author} {\bibfnamefont {G.-H.}\ \bibnamefont {Wu}},\ }\href
  {\doibase 10.1103/PhysRevD.59.096001} {\bibfield  {journal} {\bibinfo
  {journal} {Phys. Rev.}\ }\textbf {\bibinfo {volume} {D59}},\ \bibinfo {pages}
  {096001} (\bibinfo {year} {1999})},\ \Eprint
  {http://arxiv.org/abs/hep-ph/9810552} {arXiv:hep-ph/9810552 [hep-ph]}
  \BibitemShut {NoStop}%
\bibitem [{\citenamefont {Botella}\ \emph {et~al.}(2010)\citenamefont
  {Botella}, \citenamefont {Branco},\ and\ \citenamefont
  {Rebelo}}]{Botella:2009pq}%
  \BibitemOpen
  \bibfield  {author} {\bibinfo {author} {\bibfnamefont {F.~J.}\ \bibnamefont
  {Botella}}, \bibinfo {author} {\bibfnamefont {G.~C.}\ \bibnamefont {Branco}},
  \ and\ \bibinfo {author} {\bibfnamefont {M.~N.}\ \bibnamefont {Rebelo}},\
  }\href {\doibase 10.1016/j.physletb.2010.03.014} {\bibfield  {journal}
  {\bibinfo  {journal} {Phys. Lett.}\ }\textbf {\bibinfo {volume} {B687}},\
  \bibinfo {pages} {194} (\bibinfo {year} {2010})},\ \Eprint
  {http://arxiv.org/abs/0911.1753} {arXiv:0911.1753 [hep-ph]} \BibitemShut
  {NoStop}%
\bibitem [{foo(ults)}]{foot}%
  \BibitemOpen
  \href@noop {} {\bibfield  {journal} {\bibinfo  {journal} {At the one loop
  level, the electroweak gauge couplings and the $\epsilon_{ij}$ terms, induce
  quartic terms like $|\Phi_3^\dagger \Phi_{12}|^2$}\ } (\bibinfo {year} {we
  have verified that including these subdominant terms do not affect our main
  results})}\BibitemShut {NoStop}%
\bibitem [{\citenamefont {Peskin}\ and\ \citenamefont
  {Takeuchi}(1990)}]{Peskin:1990zt}%
  \BibitemOpen
  \bibfield  {author} {\bibinfo {author} {\bibfnamefont {M.~E.}\ \bibnamefont
  {Peskin}}\ and\ \bibinfo {author} {\bibfnamefont {T.}~\bibnamefont
  {Takeuchi}},\ }\href {\doibase 10.1103/PhysRevLett.65.964} {\bibfield
  {journal} {\bibinfo  {journal} {Phys. Rev. Lett.}\ }\textbf {\bibinfo
  {volume} {65}},\ \bibinfo {pages} {964} (\bibinfo {year} {1990})}\BibitemShut
  {NoStop}%
\bibitem [{\citenamefont {Peskin}\ and\ \citenamefont
  {Takeuchi}(1992)}]{Peskin:1991sw}%
  \BibitemOpen
  \bibfield  {author} {\bibinfo {author} {\bibfnamefont {M.~E.}\ \bibnamefont
  {Peskin}}\ and\ \bibinfo {author} {\bibfnamefont {T.}~\bibnamefont
  {Takeuchi}},\ }\href {\doibase 10.1103/PhysRevD.46.381} {\bibfield  {journal}
  {\bibinfo  {journal} {Phys. Rev.}\ }\textbf {\bibinfo {volume} {D46}},\
  \bibinfo {pages} {381} (\bibinfo {year} {1992})}\BibitemShut {NoStop}%
\bibitem [{\citenamefont {Olive}\ \emph {et~al.}(2014)\citenamefont {Olive}
  \emph {et~al.}}]{Agashe:2014kda}%
  \BibitemOpen
  \bibfield  {author} {\bibinfo {author} {\bibfnamefont {K.~A.}\ \bibnamefont
  {Olive}} \emph {et~al.} (\bibinfo {collaboration} {Particle Data Group}),\
  }\href {\doibase 10.1088/1674-1137/38/9/090001} {\bibfield  {journal}
  {\bibinfo  {journal} {Chin. Phys.}\ }\textbf {\bibinfo {volume} {C38}},\
  \bibinfo {pages} {090001} (\bibinfo {year} {2014})}\BibitemShut {NoStop}%
\bibitem [{foo(sent)}]{foot1}%
  \BibitemOpen
  \href@noop {} {\bibfield  {journal} {\bibinfo  {journal} {In the
  $2$-generation limit, considered here, the ETC contributions are expected to
  be real}\ } (\bibinfo {year} {only CP conserving contributions are
  present})}\BibitemShut {NoStop}%
\bibitem [{\citenamefont {Hill}\ and\ \citenamefont
  {Simmons}(2003)}]{Hill:2002ap}%
  \BibitemOpen
  \bibfield  {author} {\bibinfo {author} {\bibfnamefont {C.~T.}\ \bibnamefont
  {Hill}}\ and\ \bibinfo {author} {\bibfnamefont {E.~H.}\ \bibnamefont
  {Simmons}},\ }\href {\doibase 10.1016/S0370-1573(03)00140-6} {\bibfield
  {journal} {\bibinfo  {journal} {Phys. Rept.}\ }\textbf {\bibinfo {volume}
  {381}},\ \bibinfo {pages} {235} (\bibinfo {year} {2003})},\ \bibinfo {note}
  {[Erratum: Phys. Rept.390,553(2004)]},\ \Eprint
  {http://arxiv.org/abs/hep-ph/0203079} {arXiv:hep-ph/0203079 [hep-ph]}
  \BibitemShut {NoStop}%
\bibitem [{\citenamefont {Baak}\ \emph {et~al.}(2014)\citenamefont {Baak},
  \citenamefont {Cúth}, \citenamefont {Haller}, \citenamefont {Hoecker},
  \citenamefont {Kogler}, \citenamefont {Mönig}, \citenamefont {Schott},\ and\
  \citenamefont {Stelzer}}]{Baak:2014ora}%
  \BibitemOpen
  \bibfield  {author} {\bibinfo {author} {\bibfnamefont {M.}~\bibnamefont
  {Baak}}, \bibinfo {author} {\bibfnamefont {J.}~\bibnamefont {Cúth}}, \bibinfo
  {author} {\bibfnamefont {J.}~\bibnamefont {Haller}}, \bibinfo {author}
  {\bibfnamefont {A.}~\bibnamefont {Hoecker}}, \bibinfo {author} {\bibfnamefont
  {R.}~\bibnamefont {Kogler}}, \bibinfo {author} {\bibfnamefont
  {K.}~\bibnamefont {Mönig}}, \bibinfo {author} {\bibfnamefont
  {M.}~\bibnamefont {Schott}}, \ and\ \bibinfo {author} {\bibfnamefont
  {J.}~\bibnamefont {Stelzer}} (\bibinfo {collaboration} {Gfitter Group}),\
  }\href {\doibase 10.1140/epjc/s10052-014-3046-5} {\bibfield  {journal}
  {\bibinfo  {journal} {Eur. Phys. J.}\ }\textbf {\bibinfo {volume} {C74}},\
  \bibinfo {pages} {3046} (\bibinfo {year} {2014})},\ \Eprint
  {http://arxiv.org/abs/1407.3792} {arXiv:1407.3792 [hep-ph]} \BibitemShut
  {NoStop}%
\bibitem [{\citenamefont {Ciuchini}\ \emph {et~al.}(2013)\citenamefont
  {Ciuchini}, \citenamefont {Franco}, \citenamefont {Mishima},\ and\
  \citenamefont {Silvestrini}}]{Ciuchini:2013pca}%
  \BibitemOpen
  \bibfield  {author} {\bibinfo {author} {\bibfnamefont {M.}~\bibnamefont
  {Ciuchini}}, \bibinfo {author} {\bibfnamefont {E.}~\bibnamefont {Franco}},
  \bibinfo {author} {\bibfnamefont {S.}~\bibnamefont {Mishima}}, \ and\
  \bibinfo {author} {\bibfnamefont {L.}~\bibnamefont {Silvestrini}},\ }\href
  {\doibase 10.1007/JHEP08(2013)106} {\bibfield  {journal} {\bibinfo  {journal}
  {JHEP}\ }\textbf {\bibinfo {volume} {1308}},\ \bibinfo {pages} {106}
  (\bibinfo {year} {2013})},\ \Eprint {http://arxiv.org/abs/1306.4644}
  {arXiv:1306.4644 [hep-ph]} \BibitemShut {NoStop}%
\bibitem [{\citenamefont {Beringer}\ \emph {et~al.}(2012)\citenamefont
  {Beringer} \emph {et~al.}}]{Beringer:1900zz}%
  \BibitemOpen
  \bibfield  {author} {\bibinfo {author} {\bibfnamefont {J.}~\bibnamefont
  {Beringer}} \emph {et~al.} (\bibinfo {collaboration} {Particle Data Group}),\
  }\href {\doibase 10.1103/PhysRevD.86.010001} {\bibfield  {journal} {\bibinfo
  {journal} {Phys. Rev.}\ }\textbf {\bibinfo {volume} {D86}},\ \bibinfo {pages}
  {010001} (\bibinfo {year} {2012})}\BibitemShut {NoStop}%
\bibitem [{\citenamefont {Altmannshofer}\ \emph
  {et~al.}(2015{\natexlab{b}})\citenamefont {Altmannshofer}, \citenamefont
  {Gori}, \citenamefont {Kagan}, \citenamefont {Silvestrini},\ and\
  \citenamefont {Zupan}}]{Altmannshofer:2015esa}%
  \BibitemOpen
  \bibfield  {author} {\bibinfo {author} {\bibfnamefont {W.}~\bibnamefont
  {Altmannshofer}}, \bibinfo {author} {\bibfnamefont {S.}~\bibnamefont {Gori}},
  \bibinfo {author} {\bibfnamefont {A.~L.}\ \bibnamefont {Kagan}}, \bibinfo
  {author} {\bibfnamefont {L.}~\bibnamefont {Silvestrini}}, \ and\ \bibinfo
  {author} {\bibfnamefont {J.}~\bibnamefont {Zupan}},\ }\href@noop {} {\
  (\bibinfo {year} {2015}{\natexlab{b}})},\ \Eprint
  {http://arxiv.org/abs/1507.07927} {arXiv:1507.07927 [hep-ph]} \BibitemShut
  {NoStop}%
\end{thebibliography}
%

\end{document}